\documentstyle[epsfig]{aipproc}
\def\epm{$e^+e^-$}
\def\dec{\rightarrow}
\begin{document}
\title{B Physics \\ in the Next Millennium}

\author{Marina Artuso\thanks{Work supported by the National
Science Foundation.}}
\address{Syracuse University\\
Syracuse NY 13244
}

\maketitle
\vspace{2cm}
\centerline{\it Talk
 presented at HQ98,}
\centerline{ Workshop on HEAVY QUARKS AT FIXED TARGET,}
\centerline{ Batavia, Illinois,
October 1998}
\begin{abstract}
As we approach the turn of the century, the Standard Model is still consistent with
all our experimental observations and the path to a more complete picture of
the fundamental constituents and their interactions has yet to be clearly 
identified. Beauty flavored hadrons have provided crucial experimental 
information
on several fundamental parameters of the Standard Model
and may lead to one of the most challenging test of its validity and provide
some clues on the path towards a more complete theory. Several experiments will try to
explore this rich phenomenology in the next few years. Their physics goals and discovery
potential will be compared.
\end{abstract}

\section*{Introduction}
%
%
%
%


The investigation of $B$ meson decays has provided a wealth of information on
one of the least understood aspects of the Standard Model: the quark mixing 
that underlies the complex pattern of charge-changing transition in the quark 
sector. This pattern is summarized by a 3 x 3 unitary matrix, the 
Cabibbo-Kobayashi-Maskawa (CKM) matrix:
\begin{equation}
V_{CKM} =\left(\begin{array}{ccc} 
V_{ud} &  V_{us} & V_{ub} \\
V_{cd} &  V_{cs} & V_{cb} \\
V_{td} &  V_{ts} & V_{tb}  \end{array}\right).
\end{equation}
A commonly used approximate parameterization was originally proposed by 
Wolfenstein \cite{wolf}. 
It reflects the hierarchy between the magnitude of matrix elements 
belonging to different 
diagonals. The 3 diagonal elements and the 2 elements just above the 
diagonal
are real and positive. It is defined in first order as:
\begin{equation}
V_{CKM} =\left(\begin{array}{ccc} 
1-\lambda ^2/2&  \lambda &  A\lambda ^3(\rho -i\eta)\\
-\lambda & 1-\lambda ^2/2 & A\lambda ^2\\
A\lambda ^3(1- \rho -i\eta)&  -A\lambda ^2& 
1\end{array}\right).\end{equation}

There are
several reasons why the experimental determination of the CKM parameters
is interesting. On one hand, it is important to test that it is indeed
a unitary matrix, as dictated by the Standard Model. On the other hand, the 
complex phase that is inherent in the 3-generation CKM matrix can be an
explanation for the phenomenon of $CP$ violation, so far observed only in the 
neutral K meson system. This violation is expected to be 
responsible for baryon dominance in our world  and thus the 
understanding of its mechanisms has profound implications for our understanding
of the origin and evolution of the universe.

\begin{figure}[b!] 
\centerline{\epsfig{file=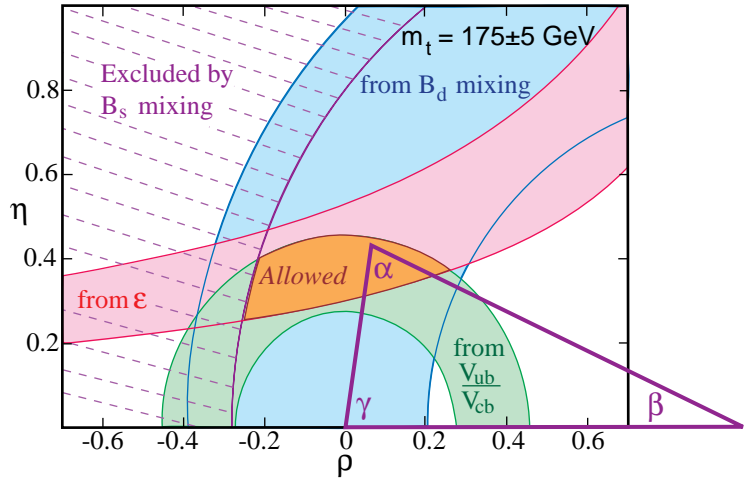,height=2.5 in}}
\vspace{10pt}
\caption{The regions in $\rho-\eta$ space (shaded) consistent
with measurements of $CP$ violation in $K_L^o$ decay ($\epsilon$), $V_{ub}/V_{cb}$
in semileptonic $B$ decay, $B_d^o$ mixing, and the excluded region from
limits on $B_s^o$ mixing. The allowed region is defined by the overlap of
the 3 permitted areas, and is where the apex of the  CKM triangle  sits. The
bands represent $\pm 1\sigma$ errors. The large width of the $B_d$ mixing band is
dominated by the uncertainty in the parameter $f_B$. Here the range  is taken as 
$240> f_B > 160$ MeV.}
\label{fig1}
\end{figure}

The $b$ 
quark provides a unique opportunity to study several CKM parameters.  
The study of 
semileptonic decays
allows the measurement of  $\left| V_{cb}\right|$ and 
$\left| V_{ub}\right|$.  $B^o-\bar{B}^o$ mixing provides information on $V_{td}$
and $V_{ts}$.  These different measurements provide independent constraints
on the `unitarity triangle' shown in  Fig. 1. A more accurate determination of the 
magnitude of $V_{ub}/V_{cb}$ and of the mixing parameters can pin down two
sides of this triangle. Note that one side has a length equal to one by 
construction. In addition, several experiments will try to get some information
on the angles $\alpha$, $\beta$ and $\gamma$. The knowledge of these angles
will answer some very fundamental questions:
\break
(1) Is the CKM phase of the three generation Standard Model the only 
source of $CP$ violation?\hfill
\break
(2) Is there new physics in the quark sector?

\section*{Future facilities for {\boldmath $B$} physics}
The next decade will see a blooming of experimental facilities planning to 
explore the $B$ decay phenomenology with increasing sensitivity to different
observables and various final states. 

\begin{table}
\caption{$B$ experiments in the near future.}
\label{near}
\begin{tabular}{lcccc}
{}& {\bf CLEO III}& {\bf BaBar-Belle} & {\bf HERA-B} & {\bf CDF-D0} \\
\tableline
{\it L}(cm$^{-2}$s$^{-1}$)& 1.7$\times 10^{33}$&3$\times 10^{33}$ & 
(Int.Rate)40 MHz & 2$\times 10^{32}$\\
$\sigma _{b\bar{b}}$ & 1.15 nb & 1.15 nb & $\approx$ 10 nb & 100 $\mu$b \\
$\sigma _{b\bar{b}}/\sigma_{had}$ & 0.25 & 0.25 & $\approx 10^{-6}$ &
$\approx 10 ^{-3}$ \\
Trigger & all $B$'s & all $B$'s & $\psi$ & high $p_t\ \mu$'s \\
Time res. & very modest & modest & good & good \\
PID & $e,\mu ,\pi , K,p$ & $e,\mu ,\pi , K,p$ &  $e, \mu , \pi ,K,p$ &
$e,\mu $\\ 
\end{tabular} 
\end{table}

Table \ref{near} summarizes the main properties of the experiments that will
take data in the near future. Among
them there are upgraded versions of previous experiments that have contributed
to our present knowledge of $B$ physics and some new facilities, HERA-B, a fixed 
target experiment at HERA, Hamburg, Germany, and the two experiments taking data 
at the new asymmetric \epm\ $B$-factories, BaBar at SLAC and Belle at KEK. 

\begin{table}
\caption{$B$ experiments starting around 2005.}
\label{far}
\begin{tabular}{lcccc}
{}& {\bf \epm\ b-factories}& {\bf ATLAS/CMS} & {\bf LHCB} & {\bf BTeV} \\
\tableline
{$L$}(cm$^{-2}$s$^{-1}$)& $10^{34}$ & 10$^{33}$(first run)  &
1.5$\times 10^{32}$  &$2\times 10^{32}$\\
$\sigma _{b\bar{b}}$ & 1.15 nb & 500$\mu$b & 500$\mu$b  & 100 $\mu$b \\
$\sigma _{b\bar{b}}/\sigma_{had}$ & 0.25 & $\approx 5\times 10^{-3}$ & 
 $\approx 5\times 10^{-3}$ & 
$\approx 10 ^{-3}$ \\
L1 Trigger & all $B$'s & high $p_t \mu$'s & medium $p_t\ \mu ,e, h$' &
 detached vertices\\
Time res. & modest & good & very good & very good \\
PID & $e,\mu ,\pi , K,p$ & $e,\mu$ &  $e, \mu , \pi ,K,p$ & $e,\mu ,\pi , K,p$\\ 
\end{tabular} 
\end{table}

A few years later, ATLAS and CMS should start taking data at the new 
LHC pp collider and they are also planning to address some of the $B$ physics 
issues discussed below. The experiments at hadronic machines discussed so 
far are pursuing $B$ physics, but they have all been optimized for their
main goal,
 high $p_t$ physics. They take advantage of the 
high cross section for $b$ production, but have not 
been designed to study $b$ decays. Two experiments have been proposed to 
exploit  the full discovery potential offered by the high cross section and 
richness of final states accessible at hadron machines: LHCb, approved to take 
data at LHC, and BTeV,
planning to take data at Fermilab. BTeV is an official R\&D project at
 Fermilab. Table \ref{far} summarizes the most
distinctive features of this next round of experiments that are expected
to take data around the year 2005.

   The 
traditional advantage of \epm\ machines operating at the $\Upsilon ({\rm 4S})$
 are their low non-$B$ background. In addition,
 the final state is composed just of a $B\bar{B}$ 
meson pair, making it easy to apply powerful kinematical constraints to
 identify 
specific final states or to reconstruct inclusive decays, like $b\dec s \gamma$
 or decays with missing 
particles like neutrinos. On the other hand, in order to measure rare decays or 
tiny $CP$ asymmetries it is necessary to collect huge data samples, posing a 
significant challenge to the accelerator physicists striving to design machines 
of ever increasing luminosity. 

Experiments taking place at hadronic facilities have the advantage of
copious production of $b$-flavored hadrons. On the other hand, their main 
challenge is the identification of the interesting $b$ events from the much more 
frequent `minimum bias' events. The key detector element in this endeavour has 
been a high resolution vertex detector, as the distinctive feature of $b$ decays 
in this environment is that their lifetime is longer than the one of the light 
quark products. CDF has been quite successful in exploiting this feature as a 
tool for $b$ physics.

\begin{figure}[b!] 
\centerline{\epsfig{file=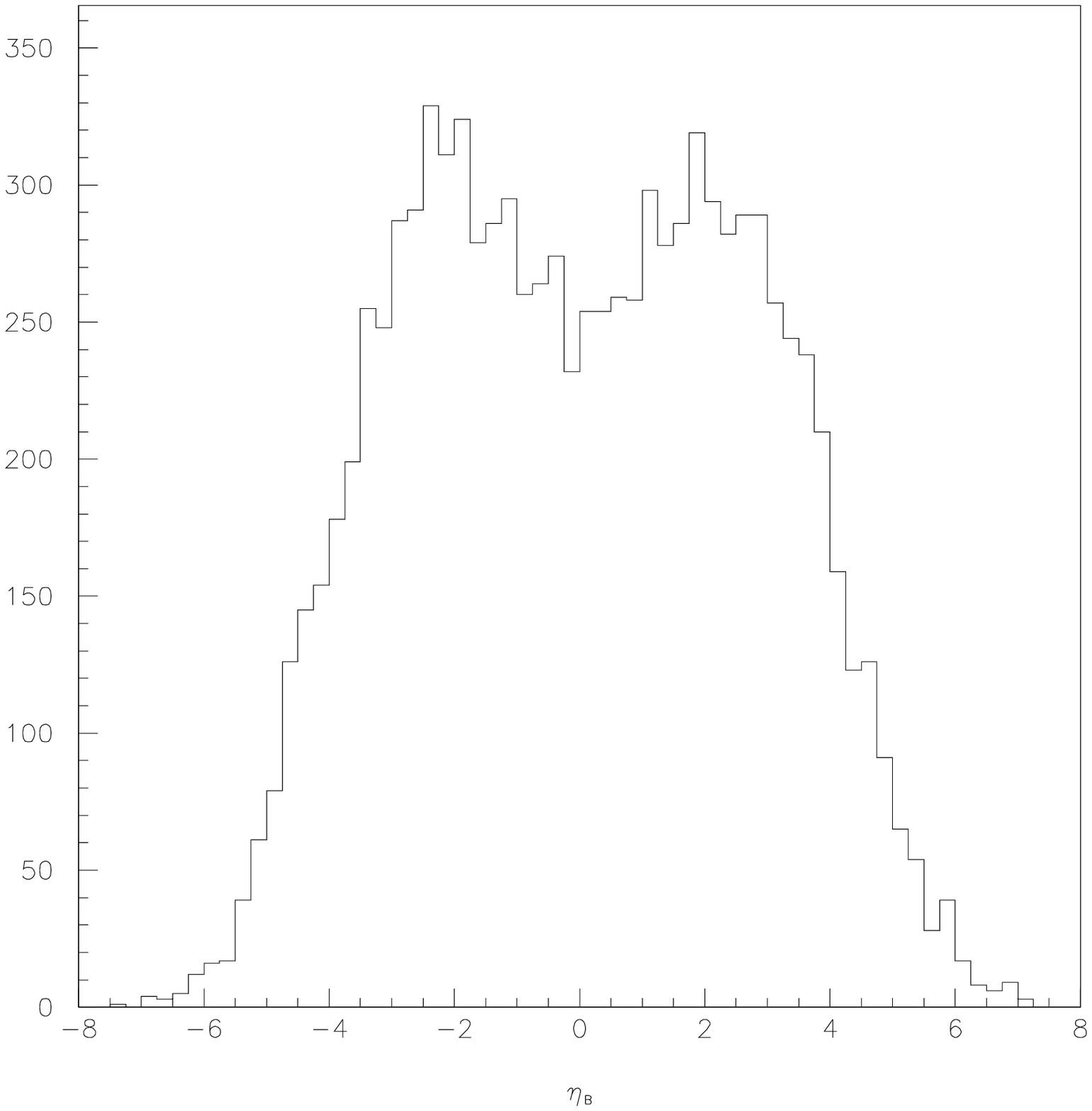, height=2.in}
\epsfig{file=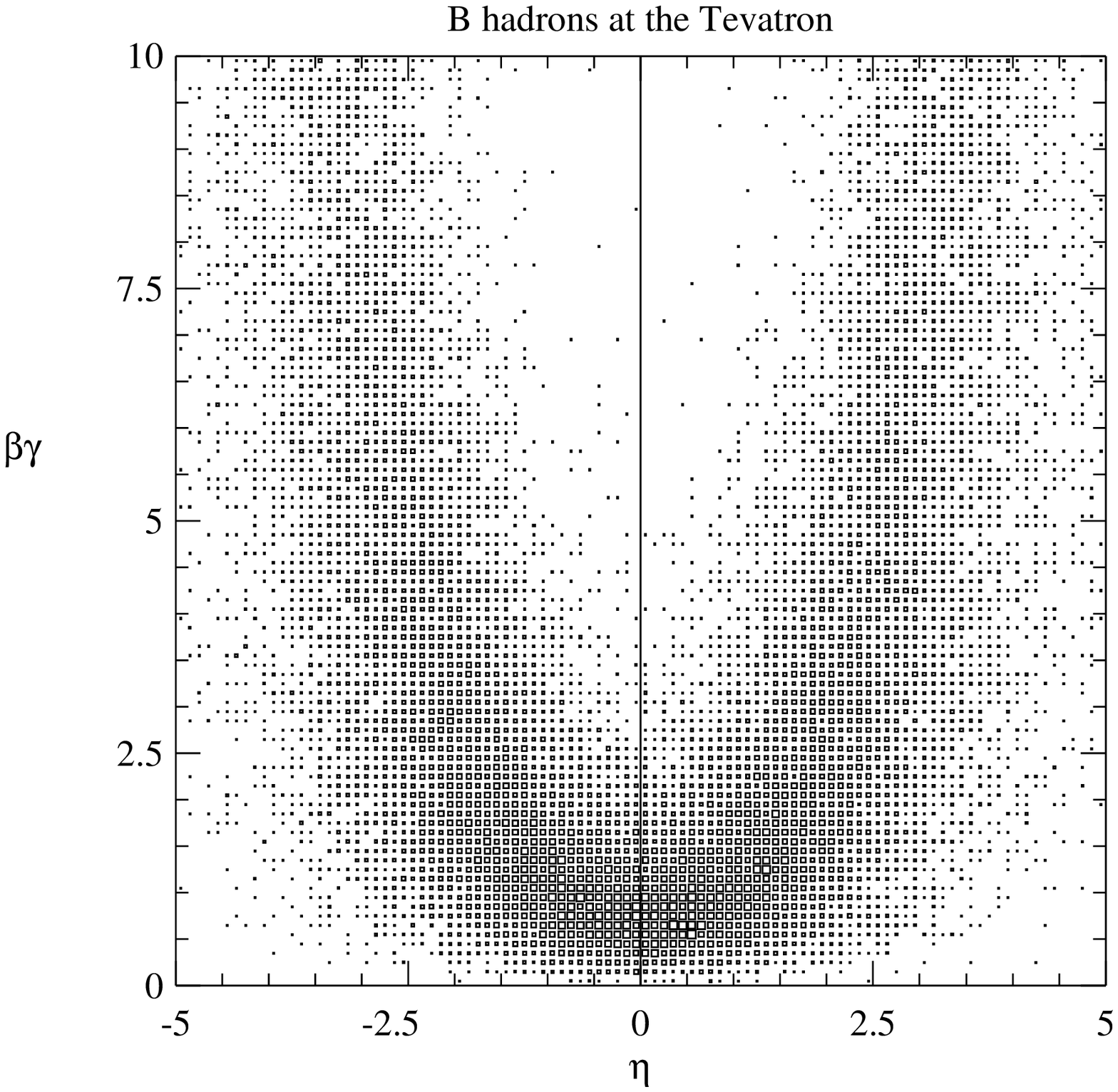,height=2.7 in}}
\vspace{2pt}
\caption{The $B$ yield plotted versus $\eta$ (left).$\beta\gamma$ of the $B$
plotted versus
$\eta$ (right). Both plots are for the Tevatron.}
\label{bprod}
\end{figure}

\begin{figure}[b!] 
\centerline{\epsfig{file=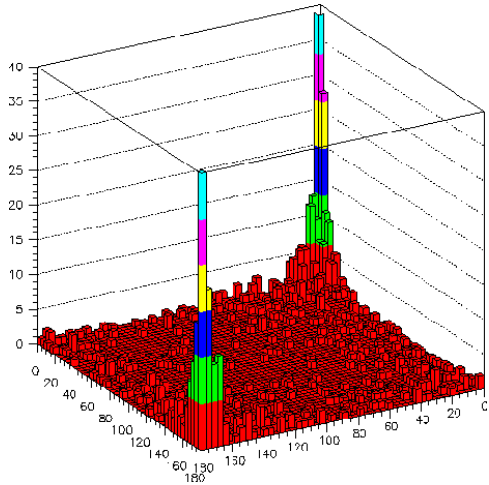,height=2.5in}}
\vspace{2pt}
\caption{The production angle (in degrees) for a hadron containing a $b$
quark plotted versus the production angle (in degrees) for a hadron containing a 
$\bar{b}$ quark.}
\label{btheta}
\end{figure}

Experiments at hadron machines designed for
$b$ physics have observed that the forward region is the most favorable to the study of 
$b$ decays. Several characteristics of hadronic $b$ 
production favor the forward direction as the region of choice for the detector 
acceptance. Fig. \ref{bprod}  shows that the $b$ quarks are produced at the 
Tevatron with a relatively  flat pseudo-rapidity distribution, where the 
pseudo-rapidity 
$\eta$ is defined as $\eta= -ln(tan(\theta /2))$ where
 $\theta$ is the 
polar angle with respect to the beam axis. Note that the more forward the $b$,
the higher its Lorentz boost is, as shown in Fig. \ref{bprod}, thus 
making it easier to identify the $b$ decays by their detached decay vertices. 
Finally, Fig. \ref{btheta}
shows the correlation in the production angles of the $b$ and $\bar{b}$ quarks. Note that 
in the forward directions the pair shows a strong correlation in their production angle. 
On the other hand, in the central region 
there isn't any correlation.
 
 BTeV and LHCb have several features in common. In
particular, both experiments take advantage of the open detector geometry in
the forward direction to include a Ring Imaging Cherenkov detector,
allowing an excellent discrimination between $\pi$'s, $p$'s and $K$'s, crucial
to some of the measurements discussed below.
 The most distinctive feature of the BTeV
experiment is a unique vertex detector, based on high resolution pixel devices
inside a dipole magnetic field, associated with fast trigger processors that 
will provide the detached vertex information at the first level trigger. This 
feature is critical to achieve high efficiency for hadronic decay modes,
like $B\rightarrow \pi ^+ \pi ^-$. In addition, it is a two arm spectrometer 
and takes advantage of the extended luminous region of the Tevatron
 ($\sigma _Z
\approx$ 30 cm) to utilize also multiple interactions per
crossing. These features more than compensate the lower $\sigma _{b\bar{b}}$
than LHCb and the two experiments are quite competitive.

\section*{A scenario for the unfolding of the {\boldmath $CP$} asymmetries}
In the Standard Model, $CP$ violation in B decays may occur whenever there are at 
least two weak  decay amplitudes with different CKM coefficients that lead to
a given final state. In the charged $B$ decay
the two decay mechanisms are provided by two competing decay diagrams, for 
instance the spectator quark decay and the so called `penguin' diagrams. On
the other hand, in neutral $B$ decays, because of $B^o - \bar{B}^o$ mixing, a
$B^o$ may decay to a final state $f$ via two paths: $B^o\rightarrow B^o 
\rightarrow f$ or  $B^o\rightarrow \bar{ B}^o 
\rightarrow f$. The phases in the second path differ from the phases in the
first because of the phase in the mixing diagram and sometimes because
of the phase difference between $B^o\rightarrow f$ and $\bar{B}^o \rightarrow
f$. In the case of charged $B$ decays, we have only direct $CP$ violation,
whereas in the case of neutral $B$ decays we can have both $CP$ violation
induced by mixing and direct $CP$ violation. In this paper we cannot summarize
all the facets of this rich phenomenology, discussed in other excellent 
reviews \cite{tony},\cite{nirquinn}. As an illustration of the challenges involved 
in the measurement, we recall that for neutral $B$ decays
to $CP$ eigenstates, only the mixing contribution is present and the time
dependent asymmetry can be expressed as:
\begin{eqnarray}
A(f_{CP})\equiv \frac{\Gamma(B^o(t)\rightarrow f_{CP})-
\Gamma(\bar{B}^o(t)\rightarrow f_{CP})}{\Gamma(B^o(t)\rightarrow f_{CP})+
\Gamma(\bar{B}^o(t)\rightarrow f_{CP})}=\nonumber \\
\chi sin(2\phi _i -\Phi _M)\frac{x}{1+x^2}
\end{eqnarray}
where $\chi =\pm 1$ is the sign of the $CP$ parity of the eigenstate, 
$\phi _i$ are the 
$CP$ violating phases related to the relevant CKM parameters, $\Phi _M$ is
the phase in the $B^o -\bar{B}^o$ mixing and $x\equiv \Delta M/\Gamma$ characterizing
 $B^o-\bar{B}^o$ mixing. The difference
$2\phi _i - \Phi _M$ is related to the quark mixing parameters: different $CP$ 
asymmetries will provide information on different angles of the unitarity 
triangle. 

In order to measure the $CP$ asymmetries there are three crucial ingredients:
adequate data samples, as often it is necessary to measure tiny differences
between final states that have a quite small branching fraction, the
ability of tagging the flavor of the initial meson and finally an adequate
suppression of backgrounds. 

\begin{table}
\caption{Summary of $\epsilon D^2$ available at different facilities: forward
and central refer to the detector geometry at a hadron collider.}
\label{btag}
\begin{tabular}{lccc}
{\bf Method}& {\bf Forward}\cite{btev} & {\bf Central} \cite{cdf}&
 {\epm b-factories}\cite{babarp}\\
\tableline
$K^{\pm}$ & 5\% & 0\%-2.4\% & 9.6\%\\
$\mu ^{\pm}$&  1.6 \%& 1.0 & 5.4\% \\
$e^{\pm}$ & 1.0\% &  0.7\% & 8.4\%\\
SST & $> 2\%$ & 2\%& - \\
Jet Charge & 6.5 \% & 3\% & -\\
\end{tabular}
\end{table}

The development of a variety of flavor tagging techniques in the different experimental
configurations has been one of the most active area of investigation
towards the development of the physics analysis tools at different
facilities. Some of the tagging techniques, like the charge of the $\mu$'s
produced in semi-leptonic decays, are common to most experiments,  others
are environment-specific. For example, \epm\ b-factories can try to
take advantage of the low momentum leptons produced in the cascade
$B\dec D\dec K^{(*)} \ell \nu$. On the other hand, hadronic machines
can take advantage of same side tagging, exploiting
the correlation between the flavor of the $B$ hadron and the charge of
the pion produced in close association with it. Table \ref{btag} illustrates
a comparison between the tagging efficiency of different approaches.

\begin{table}
\caption{Prospects for $\sin{2\beta}$ with 1 year of running at the nominal luminosity}
\label{sbeta}
\begin{tabular}{lll}
{\bf Experiment} &{\bf $\delta \sin{2\beta }$} & {\bf Remarks}\\
\hline
BaBar & $\pm$0.09 &  using $\psi K^0$ \cite{babarp} assuming $sin(2\beta)$=0.7\\
Belle & $\pm$0.11 & using $\psi K^0$ \cite{kek} \\
HERA-B & $\pm$0.13 & using $\psi K_S$ \cite{iris}\\
CDF & $\pm$0.09&  using $\psi K_S$ \cite{cdf}\\
D0 & $\pm$0.11 & using $\psi K_S$ \cite{d0}\\
BTeV &$\pm$0.013 & using $\psi K_S$ \cite{btev}\\
LHCb & $\pm$0.017-0.011 &  using $\psi K_S$ and $sin(2\beta)$=0-0.866 \cite{lhcbt}\\
ATLAS &$\pm$0.018 &  using $\psi K_S$ \cite{earola} \\
CMS & $\pm$0.058 &  using $\psi K_S$\cite{cms} \\
\hline
\end{tabular}
\end{table}

Table \ref{sbeta} shows the prospects for the $\sin{2\beta }$ measurement by the
different experiments. Note that the projections are taken from proposals and
simulation studies performed by the proponents of the various experiments and
do not necessarily share the same level of realism. The data shown here and
in the following discussion should be taken as indicating some trends rather 
than as a detailed quantitative comparison. 

\begin{table}
\label{salpha}
\caption{Prospects for $\sin{2\alpha}$ with 1 year of running at nominal luminosity}
\begin{tabular}{lll}
{\bf Experiment} &{\bf $\delta \sin{2\alpha }$} & {\bf Remarks}\\
\hline
BaBar & $\pm$0.29 & using $\pi ^+ \pi ^-$ \cite{babarp}\\
Belle & $\pm$0.27 & using $\pi ^+\pi ^-$(assuming no penguin)\cite{kek} \\
CDF & $\pm$0.22&  using $\pi^+ \pi^-$ and assuming PID=TOF+dE/dx \cite{cdf}\\
BTeV & $\pm$0.026 & using $\pi ^+ \pi^-$ (no penguin) \cite{btev}\\
LHCb & $\pm$0.05& using $\pi ^+ \pi^-$ (no penguin) \cite{lhcbt}\\ 
ATLAS & $\pm 0.1 \sin(2\alpha ) \oplus 0.011$ &  using $\pi ^+ \pi^-$ (no penguin,) \cite{earola}\\  
CMS & $\pm$0.067 &  using $\pi ^+ \pi^-$ (no penguin, statistical error only)\cite{cms} \\  
\hline
\end{tabular}
\end{table}

\begin{figure}[b!] 
\centerline{\epsfig{file=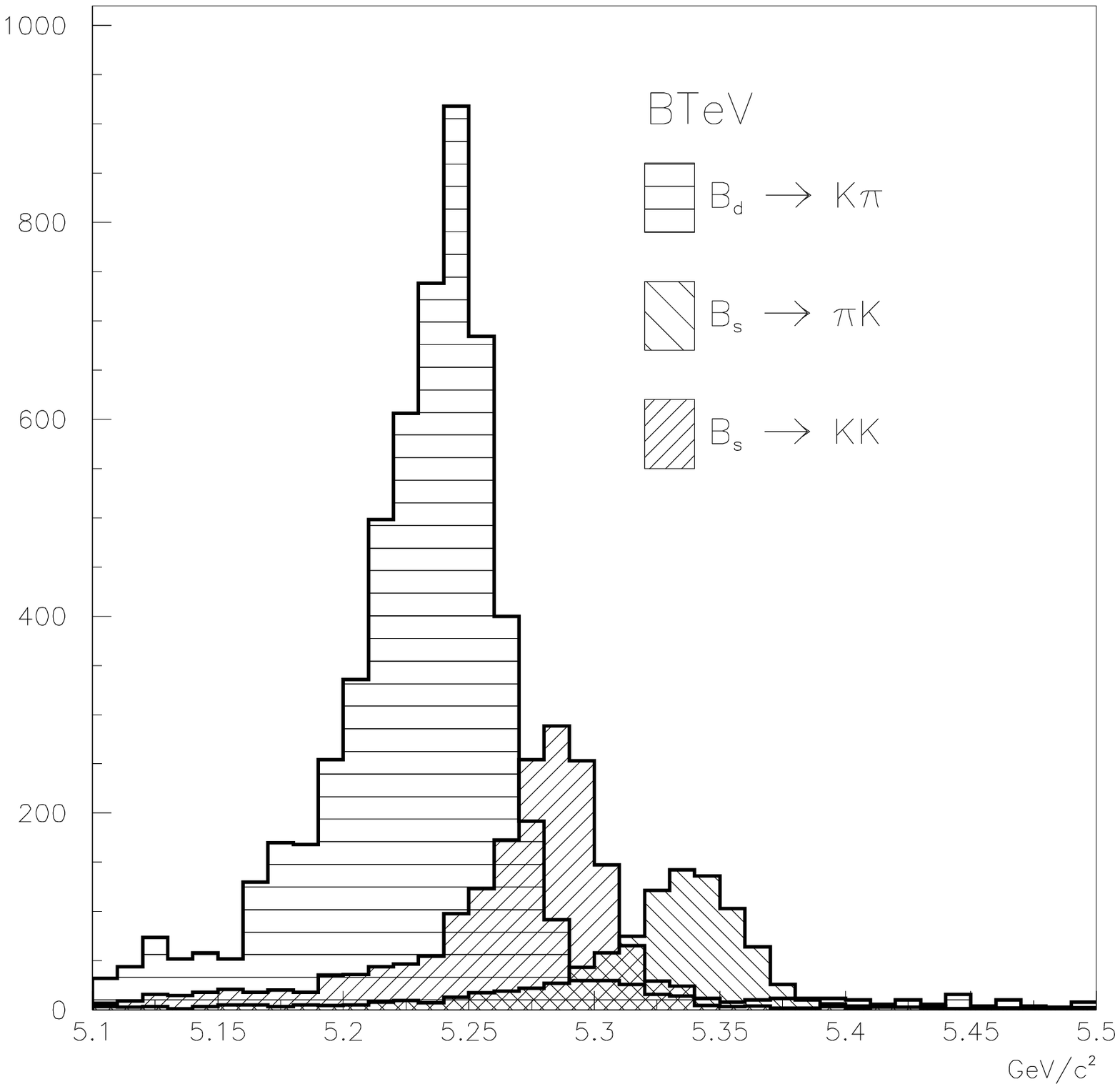, height=2.5in}
\epsfig{file=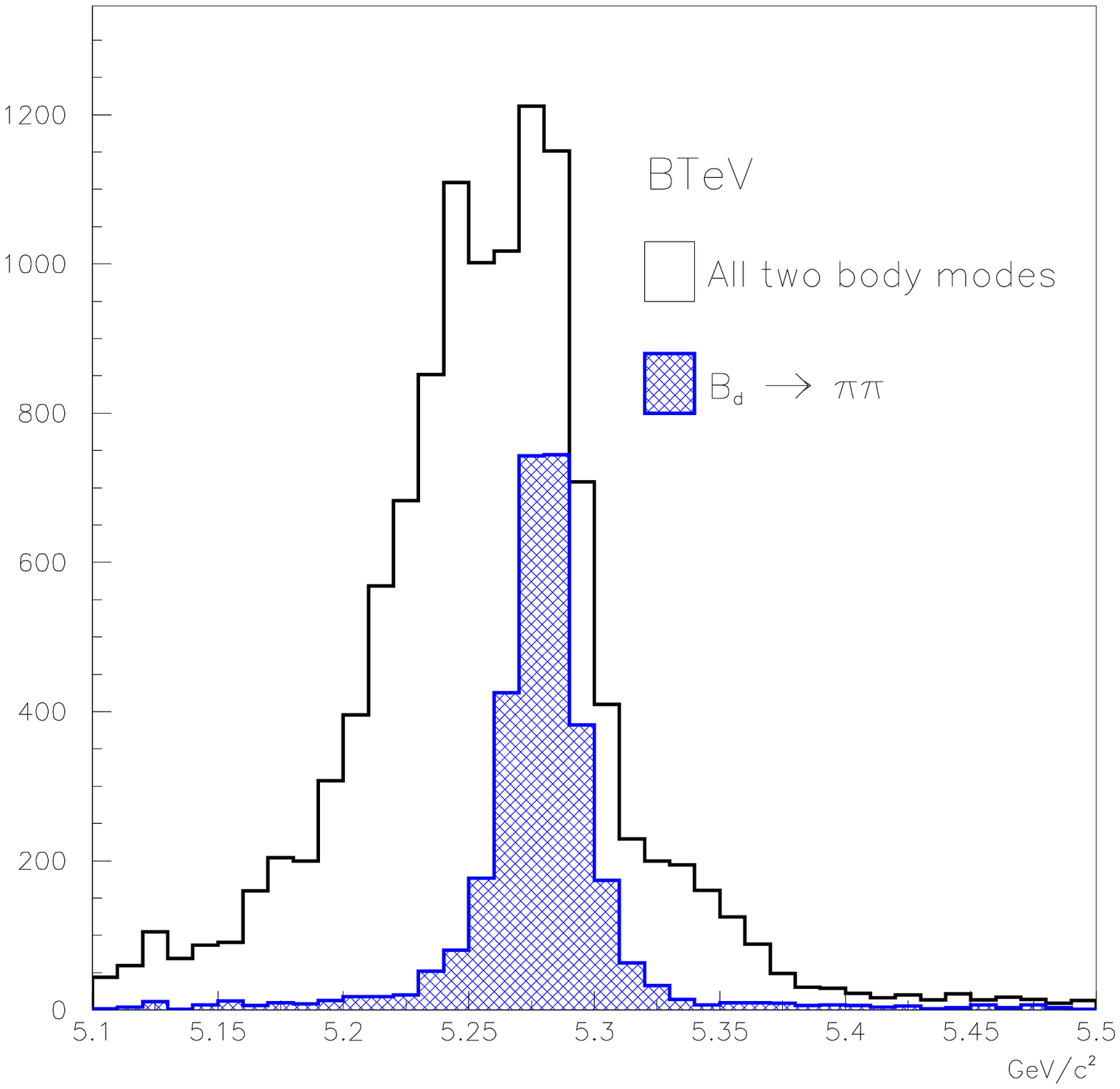,height=2.5in}}
\vspace{2pt}
\caption{Invariant mass distribution for all the $B\dec h^+h^-$ final 
states, where $h$ denotes either a $\pi$ or a $K$, and the mass is computed
assuming that both the particles are $\pi$'s. The plot on the left shows 
all the individual background contributions and the plot on the right shows
the sum of all the channels, properly normalized (see text). The plot refers
to the Tevatron.}
\label{pipipid}
\end{figure}

The determination of the angle $\sin{2\alpha}$ is a much more complex issue.
The `golden mode' in this case used to be the decay $B^0\rightarrow
\pi ^+ \pi^-$. However `penguin pollution'\cite{ppoll} complicates
 the relationship between the measured asymmetries and
$\sin{2\alpha}$. In addition to the spectator diagram, that would provide a
contribution to $A_{CP}$ according to the formulation discussed above,the
 penguin diagram adds a term proportional to $cos({xt})$. The
fraction of penguin contribution needs to be known to extract 
$\alpha$\cite{duni}. Note that the different experiments
simulate the effects of penguin pollution with very
different degrees of approximation. In general, the label `no
penguin' refers to simulations that assume that the penguin pollution is
negligible.  Moreover,
 recent CLEO results \cite{gaidarev} have shown that
the branching fraction for this decay may be quite smaller than anticipated, its
upper limit at 90\% C.L. being $0.8\times 10^{-5}$,
thus seriously affecting the prospects of \epm\ b-factories. Lastly, a state of the art
hadron identification system is necessary to single out this final state from
other two-body decay modes of the $B_d$ and $B_s^0$ mesons, as illustrated by
Fig. \ref{pipipid}. The data are taken from the BTeV simulation studies but 
apply to all the experiments taking data at hadronic facilities, thus showing
that the excellent particle identification featured by LHCb and BTeV are
crucial to make a reliable measurement. The normalization between the different
decay modes is obtained assuming ${\cal B}(B\dec K^+\pi^-) = 1.5\times 10^{-5}$  and
${\cal B}(B\dec \pi^+ \pi ^-) = 0.75\times 10 ^{-5}$, according to the 
recent CLEO
results \cite{gaidarev}, and ${\cal B}(B_s\dec K^+ K^-)={\cal B}( B_d
\dec K^+ \pi^-)$ and
 ${\cal B}(B_s\dec K^+ \pi^-)= {\cal B}(B_d\dec \pi^+ \pi^-)$.

The determination of the angle $\gamma$ introduces new challenges. 
In principle, decay modes of the $B_S$ meson to $CP$ eigenstates, like
$\rho K_S $, could be used. However the same `penguin pollution' problems 
alluded to in the discussion of $\sin{2\alpha }$ are present here and they
are even more difficult to be disentangled than in the previous case because
of the  vector-pseudoscalar nature of the final state. 
An alternative approach can be used to extract the angle $\gamma $
from the decays $B_S\rightarrow D_S^{\pm}K^{\mp}$, where a time-dependent $CP$ 
violation can result from the interference between the direct decay and the 
mixing induced decays \cite{twentione}. BTeV and LHCb have studied the
possibility of extracting the angle $\gamma$ with this approach. They project 
 errors of $\pm 8^{\circ}$
and $\pm 10^{\circ }$, respectively.

Another method for extracting $\gamma$ has been proposed by Atwood, Dunietz 
and Soni\cite{soni}, who refined a method suggested originally by Gronau
and Wyler \cite{gw}. A large $CP$ asymmetry can result from the interference
between the decays $B^-\rightarrow K^- D^o$,$D^o\rightarrow f$,   where $f$
is a doubly-suppressed Cabibbo decay of the $D^o$ 
(for example, $f= K^+\pi ^-$, and $B^-\dec K^- \bar{D}^o$, $D^o\rightarrow
f$, ). Since $B^- \dec K^- \bar{D}^o$ is color-suppressed and $B^-\dec
K^- D^o$ is color allowed, the overall amplitude for the two decays are
expected to be approximately equal in magnitude. The weak phase between them
is $\gamma$. The subtleties of extracting the CKM angle from the measurements
of two different states are discussed in Ref. \cite{btev}. 

Finally, Gronau and Rosner \cite{gronau} originally proposed a method based on
the study of the
decays $B\dec K\pi$. The use of these decay modes is complicated by rescattering
processes and $SU(3)$ breaking effects, as pointed out 
in several subsequent papers. However the intense theoretical
effort to understand these decay modes of the neutral and charged $B$ meson
can ultimately 
provide a good strategy to extract the angle $\gamma$. A recent analysis
by A. Buras and G. Fleischer\cite{burasfl} examines all the hadronic effects in great 
detail and gives some promising strategies to use this approach more 
effectively.

A complementary constraint to the unitarity triangle is provided by the measurement of
$B_s\bar{B}_s$ mixing, using the ratio:
\begin{equation}
\left|\frac{V_{td}}{V_{ts}}\right|^2=\xi ^2\frac{m_{B_s}}{m_{B_d}}
\times \frac{\Delta m_d}
{\Delta m_s}
\end{equation}
where $\xi = f_{B_s}\sqrt{{\cal B}_{B_s}}/ f_{B}\sqrt{{\cal B}_{B}} =1.15\pm 0.05$
is the SU(3) breaking term, estimated from lattice and QCD
sum rules.  The time resolution is crucial in this measurement. The projected proper time
resolution for BTeV is about 30 fs\cite{btev}, and for LHCb is 43 fs\cite{lhcbt}, whereas for CDF is 60 fs
(possibly down to 46 fs with an additional silicon layer) \cite{cdfprop} and
for ATLAS is 64 fs.  BTeV expects to be able to measure $\Delta m_s$ at least up to 
51 ps$^{-1}$ within a reasonable time scale. LHCb expects to measure  $\Delta m_s$ 
with a statistical significance of at least 5 $\sigma$ if the true value of
$\Delta m_s\le 48$
ps$^{-1}$ or exclude values of $\Delta M_s$ at 95\% C.L. up to 58 ps$^{-1}$ (corresponding
to a value of $x_s\equiv \Delta M_s/\Gamma_s =91)$. For comparison, CDF claims to be able
to make this measurement if $x_s \le 20$, D0 claims to be able to make this measurement
if $x_s \le 16$ and ATLAS and CMS if $x_s \le 38$. Note that the extraction of $\left|V_{td}/V_{ts}\right|$ with this method minimizes the
theoretical uncertainty and therefore this observation will have a significant impact 
on our understanding of the CKM matrix. 

\section*{Conclusions}
This paper illustrates how different experiments will contribute to a
 precision determination of the 
angles and sides involved in the unitarity triangle and thus provide a crucial
 test of the validity of
the CKM picture of quark mixing and $CP$ violation. In addition, they will
 perform detailed studies of
 rare $B$ decays, thus providing 
complementary tests of the Standard Model and useful constraints on more 
exotic models, like SUSY or 
a more complex Higgs sector. In the next year, experiments at asymmetric
$e^+e^-$ $B$ 
factories, Babar and Belle, will start collecting data. They are likely to
make the first significant measurements of $\sin 2\beta$. In the same time period,
the symmetric $B$ factory experiment CLEO will start with its III upgraded
version. If it proves easier to make luminosity with a single ring symmetric
energy machine, they may be the first to see direct CP violation and will
continue their measurements of rare $B$ decays that have already provided
quite interesting results \cite{gaidarev}.
With the start of Tevatron Run II, CDF and D0 will try to beat the 
the $e^+e^-$ machines to the first measurements of $\sin 2\beta$.
HERA-B will also enter the race. Ultimately, crucial measurements on $B_s$
mixing, $\sin 2\alpha$, $\gamma$ and very rare $B$ decays are likely to
be made at
forward experiments at hadron machines, LHCb or BTeV, where the
$B$ rates are large, the vertexing is accurate and the particle
identification is excellent.  These
measurements are  crucial to a
 complete and accurate 
picture of this complex phenomenology.

\section*{Acknowledgements}
The author would like to thank J. Butler and H. Cheung, as they distinguished
themselves among the organizers for their indefatigable dedication to the 
rich scientific program and the smooth running of this very enjoyable conference.
In addition, many thanks are due to S. Stone, I. Bigi and A.I. Sanda for 
interesting
discussions. Lastly I would like to thank Julia Stone for many pleasant
interludes during the writing of this manuscript.



\end{document}